%% file: main.tex
\documentclass[journal]{IEEEtran}
\usepackage{amsmath,amsfonts}
\usepackage{algpseudocode}
\usepackage{algorithm}
\usepackage{amsmath,amsfonts}
\usepackage{algpseudocode}
\usepackage{algorithm}
\usepackage{array}
\usepackage[caption=false,font=normalsize,labelfont=sf,textfont=sf]{subfig}
\usepackage{textcomp}
\usepackage{stfloats}
\usepackage{url}
\usepackage{verbatim}
\usepackage{color,soul}
\usepackage{graphicx}
\usepackage{hyperref}
\usepackage{cite}
\usepackage[nolist]{acronym}

\input{acronyms.tex}

\DeclareMathOperator{\MSE}{MSE}
\DeclareMathOperator{\SNR}{SNR}
\DeclareMathOperator{\dB}{dB}
\DeclareMathOperator{\SRP}{SRP}

\begin{document}
\title{RIS-Enabled Passive Radar towards Target Localization}
\author{Ahmad Bazzi and Marwa Chafii 
\thanks{Ahmad Bazzi is with the Engineering Division, New York University (NYU) Abu Dhabi, 129188, UAE
(email: \href{ahmad.bazzi@nyu.edu}{ahmad.bazzi@nyu.edu}).

Marwa Chafii is with Engineering Division, New York University (NYU) Abu Dhabi, 129188, UAE and NYU WIRELESS,
NYU Tandon School of Engineering, Brooklyn, 11201, NY, USA (email: \href{marwa.chafii@nyu.edu}{marwa.chafii@nyu.edu}).}
\thanks{Manuscript received xxx}}
\markboth{Journal of \LaTeX\ Class Files,~Vol.~14, No.~8, August~2021}%
{Shell \MakeLowercase{\textit{et al.}}: A Sample Article Using IEEEtran.cls for IEEE Journals}


\maketitle

\begin{abstract}
In this paper, we study a communication-centric integrated sensing and communication (ISAC) approach, where an access point (AP) communicates with users, while a passive radar (PR) is present in the environment. We investigate the deployment of a reconfigurable intelligent surface (RIS) to enable the PR to localize a target. We derive an optimization problem for updating the phase shifters of the RIS per epoch. Due to the limited information at the PR, such as unknown payload information and unknown number of targets in the scene, we propose two methods capable of performing joint angle of arrival estimation and detection of the targets. We demonstrate the superior performance of the methods onto the proposed setting through numerical simulations, in comparison to a no-RIS baseline scheme.
\end{abstract}

\section{Introduction}

\IEEEPARstart{R}esearch has begun to piece together a speculative vision of 6G by studying potential future services and applications, identifying market demands, and finding disruptive technologies \cite{chafii2022ten}. A vast range of services must be supported by 6G, including blockchain, haptic telemedicine, VR/AR remote services, holographic teleportation, and huge eXtended reality (XR) capabilities. There is no doubt that 6G has advantages beyond connectivity. It will combine new functions like sensing and computing, opening up new services and employing better environmental data for artificial intelligence and machine learning. However, as \cite{saad2019vision} emphasizes, such bandwidth-hungry applications need for a $10^3 \times$ increase in capacity. Furthermore, it is anticipated that by $2030$, mobile data use per mobile broadband subscriber would increase by a factor of 50, from $5.3$GB in $2020$ to $257$GB in $2030$. Nevertheless, there are several enabling technologies (such as sub-$6$ GHz and mmWave frequencies) and \ac{RIS}. 

By ingeniously reconfiguring the wireless propagation environment, \ac{RISs} have drawn a lot of attention for their potential to increase the capacity and coverage of wireless sensor networks. A \ac{RIS} \cite{renzo2019smart} is composed of an array of reflecting elements capable of reconfiguring incident signals, thanks to its two-dimensional material structure with programmable macroscopic physical characteristics. Since the \ac{RIS} can reconfigure the \ac{EM} incident wave, the wireless channel between different nodes of the wireless network could be altered. In fact, \ac{RIS}'s relatively low energy consumption is one of its most appealing features because it allows for the amplification and transmission of incoming signals without the use of a power amplifier. However, because of the inexpensive cost of \ac{RIS}, it may be unusually large and sprawl the entire wall of a building \cite{han2022localization}. Instead, phase shifts applied by each reflecting element are properly designed in order to combine each reflected signal in a useful manner. In the literature, they are also referred to as \ac{LISA} \cite{decarli2021communication,hou2021mimo} or \ac{IRSs} \cite{cheng2021downlink,wu2019towards}. Recently, \ac{RISs} have been used to aid many profound applications in wireless communications, such as beamforming \cite{lin2020reconfigurable,pei2021ris}, security \cite{xu2022ris,alexandropoulos2021safeguarding}, \ac{OFDM} \cite{pradhan2020reconfigurable,lin2020adaptive}, millimeter-wave channel estimation \cite{zhou2022channel,ardah2021trice,liu2021admm,he2021channel} just to name a few. There is a handful of significant advantages that could be leveraged from \ac{RIS} deployment \cite{liu2021reconfigurable} in various wireless communication scenarios, such as their capability of improving the spectral and energy efficiency \cite{huang2019reconfigurable,huang2018energy}, their ease of deployment and environment friendliness, compatibility with already existing standards \cite{zhou2020spectral}, and secure wireless communications \cite{cui2019secure}. For more advantages in terms of security, energy efficiency and coverage, the interested reader is referred to \cite{liaskos2018using}.

Within the last $80$ years, passive radar, which indicates the localization of a target by radar data without the use of own controlled emissions, had been thoroughly investigated \cite{dawidowicz2012detection,finden2015analysis,krysik2014doppler,palmer2012dvb}. It is highlighted in \cite{strinati2021wireless}, that \ac{RIS} is foreseen to be a key enabler for advanced localization and sensing by \ac{PR}. The deployment of passive radars possess exceptional advantages, due to their low cost \cite{kuschel2019tutorial} in terms of procurement, operation and maintenance, as well as its ability in covert operation. In \ac{PR}, it is crucial to locate targets with minimal information. Indeed, the passive radar, unlike traditional radar, is protocol-agnostic. That is, the \ac{PR} has no knowledge about transmit waveform information. On the other hand, \ac{PR} sometimes fail to perform target localization due to lack of aid from active components.



A challenging, yet interesting, feature of 6G is \ac{ISAC}, where the designed waveforms convey communication, as well as sensing information \cite{chafii2022ten}. This allows for shared spectrum, power efficiency, in addition to hardware efficiency. \ac{ISAC} can be split into three broad categories, where the first aims at a \textit{joint design}, offering trade-offs between sensing and communications. For example, the work in \cite{bazzi2022outage} derives beamforming matrices for \ac{ISAC} systems, under practical assumptions, such as imperfect channel state information, while \cite{xu2022robust} considers a joint design for variable length time snapshots. Another category is \textit{radar-centric} \ac{ISAC}, where the main goal is to embed communication information onto a radar waveform, such as a chirp \cite{9828505}. The last category is \textit{communication-centric} \ac{ISAC}, where one simply performs sensing using a known communication waveform, such as \ac{OFDM}. The priority in this case is communications, and sensing comes as an additional feature. This is of advantage in scenarios where the communication infrastructure is established and obeys a certain protocol (ex. 802.11be/ax/ac). In this paper, we focus on a communication-centric \ac{ISAC} approach, where the AP communicates with communication users and a \ac{PR} is deployed to localize the targets, with the aid of \ac{RIS}. An optimization framework is given to adjust the phase shifts at the \ac{RIS}, and two algorithms are presented that allow simultaneous location estimation and detection of targets.

\subsection{Existing work}

Channel estimation may be required between \ac{BS} and \ac{UE} in order to design the  reflective matrix, i.e the phase shifters at the \ac{RIS}. For example, to design the RIS phase shifts, the approach in \cite{taha2021enabling} adopts \ac{DL} and \ac{CS} techniques. This paper does not require channel estimation for target localization.  \cite{he2020adaptive} proposes an adaptive \ac{RIS} phase shifter based on hierarchical codebooks with feedback from the \ac{UE}. Even though the methodology followed in \cite{he2020adaptive} performs localization, the fundamental difference lies in the scenario itself. In particular, the approach does not make use of \ac{PR} and it does not leverage the multi-path bounces off the targets to perform localization. Likewise, the work in \cite{fascista2021ris} also performs localization with the help of an \ac{RIS} situated in the scene, but in the absence of \ac{PR}. Furthermore, even though the paper in \cite{elzanaty2021reconfigurable} presents an \ac{RIS}-assisted approach for localization, the model does not comprise of reflections resulting from targets to be localized. In particular, the received signal in \cite{elzanaty2021reconfigurable} is the direct path and the reflection resulting from the \ac{RIS}. In contrast to our proposed model, where the \ac{PR} performs spatial beamforming towards the \ac{RIS} and focuses on the reflections bouncing off the targets towards the \ac{RIS}, which in turn mirrors these bounces towards the \ac{PR}. Meanwhile, the authors in \cite{zhang2021metalocalization} aim at optimizing the \ac{RIS} phase shift coefficients to maximize the differences of \ac{RSS} values as the location changes. A similar approach utilizing \ac{RSS} information appears in \cite{zhang2020towards}. As opposed to \cite{zhang2021metalocalization,zhang2020towards}, we consider the contributions that bounce towards the \ac{RIS}, then reflected towards the \ac{PR}. In addition, our work does not utilize the \ac{RSS} to perform target localization, as it relies on the \ac{AoA} information of the different reflections.


\subsection{Contributions and Insights}
This work focuses on deploying a \ac{PR} and an \ac{RIS} in an existing communication scenario. The \ac{RIS} would then aid the \ac{PR} to perform target localization without any knowledge of the standard used. Therefore, no assumption on the transmit signal structure is required to perform target localization.   \\\\
To that purpose, we have summarized our contributions below.
\begin{itemize}
	\item \textbf{A new architecture towards target localization}. This work describes a new \ac{RIS}-aided setting dedicated for target localization, which is well-suited for communication-centric \ac{ISAC} systems. More specifically, a \ac{PR} and an \ac{RIS} are considered, where the \ac{RIS} "mirrors" the reflections resulting from targets in the scene towards the \ac{PR}.
	\item \textbf{RIS matrix optimization}. Upon "mirroring" back the reflections towards the \ac{PR}, the \ac{RIS} configures suitable phase shifters that could properly reflect the reflections caused only by the targets to localize.
	\item \textbf{Localization without knowledge of number of targets}. The \ac{PR} being unaware of the dynamics of the environment, we believe that joint estimation and detection of the targets is an important task that should be carried out at the \ac{PR}, with minor information about the signaling being used. Motivated by this task, we propose two methods for target localization without the need to resort to techniques that require source number enumeration, such as Akaike's information criterion (AIC) \cite{akaike1998information} and minimum description length (MDL) \cite{rissanen1978modeling}. The proposed methods are based on \ac{NLMS} filtering. Therefore, the proposed methods perform joint estimation and detection of targets.
\end{itemize}
Furthermore, we unveil some important insights, i.e.
\begin{itemize}
	\item A $20\dB$ gain in terms of required \ac{SNR} to attain an \ac{MSE} of $0.2$ is achieved when utilizing the \ac{RIS} aided target localization approach, with $16$ reflective elements.  
	\item Since the proposed methods perform simultaneous detection of number of targets, we also study the probability of detection. In this respect, a $16\dB$ gain in terms of required \ac{SNR} for a probability of detection equal to $0.9$ is attained with $16$ reflective  elements, as compared to a no-\ac{RIS} approach. A supplemental $4\dB$ gain is achieved, when doubling the number of reflective elements. 
	\item With respect to \ac{SRP}, we observe gains in the order of $18 \dB$ when localizing via the proposed \ac{RIS}-aided architecture.
\end{itemize}

\subsection{Organization and Notations}
The detailed structure of this paper is given as follows,
\begin{itemize}
	\item Section \ref{sec:system-model} presents the system model which includes the AP, targets and an \ac{RIS} in the scene. The section describes how the \ac{PR} senses the reflections mirrored by the \ac{RIS} and mathematically formulates the problem at hand.   
	\item Section \ref{sec:reflection-matrix-optimization} demonstrates the reflection matrix optimization technique. More specifically, it formulates and solves phase shifter components per epoch, relative to the \ac{RIS}. 
	\item Section \ref{sec:beamforming-towards-RIS} presents the beamforming step at \ac{PR} level. In particular, the \ac{PR} should look towards the \ac{RIS} in order to properly acquire all the reflections caused by the $K$ targets. This section is dedicated towards adjusting the \ac{PR}'s beamforming weights.
	\item Section \ref{sec:target-localization} presents the two methods for target localization performed at the \ac{PR}. The methods are batch and sequential versions of an \ac{NLMS} filter estimate devoted for target localization through reflections. 
	\item Section \ref{sec:simulations} showcases our simulation results and the superiority of the setting, when compared to the baseline approach, i.e. when no \ac{RIS} is employed. We perform target localization using both introduced methods and study the performance via different metrics, such as the \ac{MSE}, \ac{SRP}, and probability of detection.
	\item Section \ref{sec:conclusion} concludes the paper and provides future insights and ideas.
\end{itemize}

\textbf{Notation}: : Upper-case and lower-case boldface letters denote matrices and vectors, respectively. $(.)^T$, $(.)^*$ and $(.)^H$ represent the transpose, the conjugate and the transpose-conjugate operators. The statistical expectation is denoted as $\mathbb{E}\lbrace \rbrace$. For any complex number $z \in \mathbb{C}$, the magnitude is denoted as $\vert z \vert$, its angle is $ \angle z$. The $\ell_2$ norm of a vector $\pmb{x}$ is denoted as $\Vert \pmb{x} \Vert$. The matrix $\pmb{I}_N$ is the identity matrix of size $N \times N$. The zero-vector is $\pmb{0}$. For matrix indexing, the $(i,j)^{th}$ entry of matrix $\pmb{A}$ is denoted by $\pmb{A}_{i,j}$. The sub-matrix of $\pmb{A}$ ranging from rows $i_1$ to $i_2>i_1$ and columns $j_1$ to $j_2>j_1$ is denoted as $\pmb{A}_{i_1:i_2,j_1:j_2}$. 

\section{System Model}
\label{sec:system-model}
\begin{figure}[!t]
\centering
\includegraphics[width=3.5in]{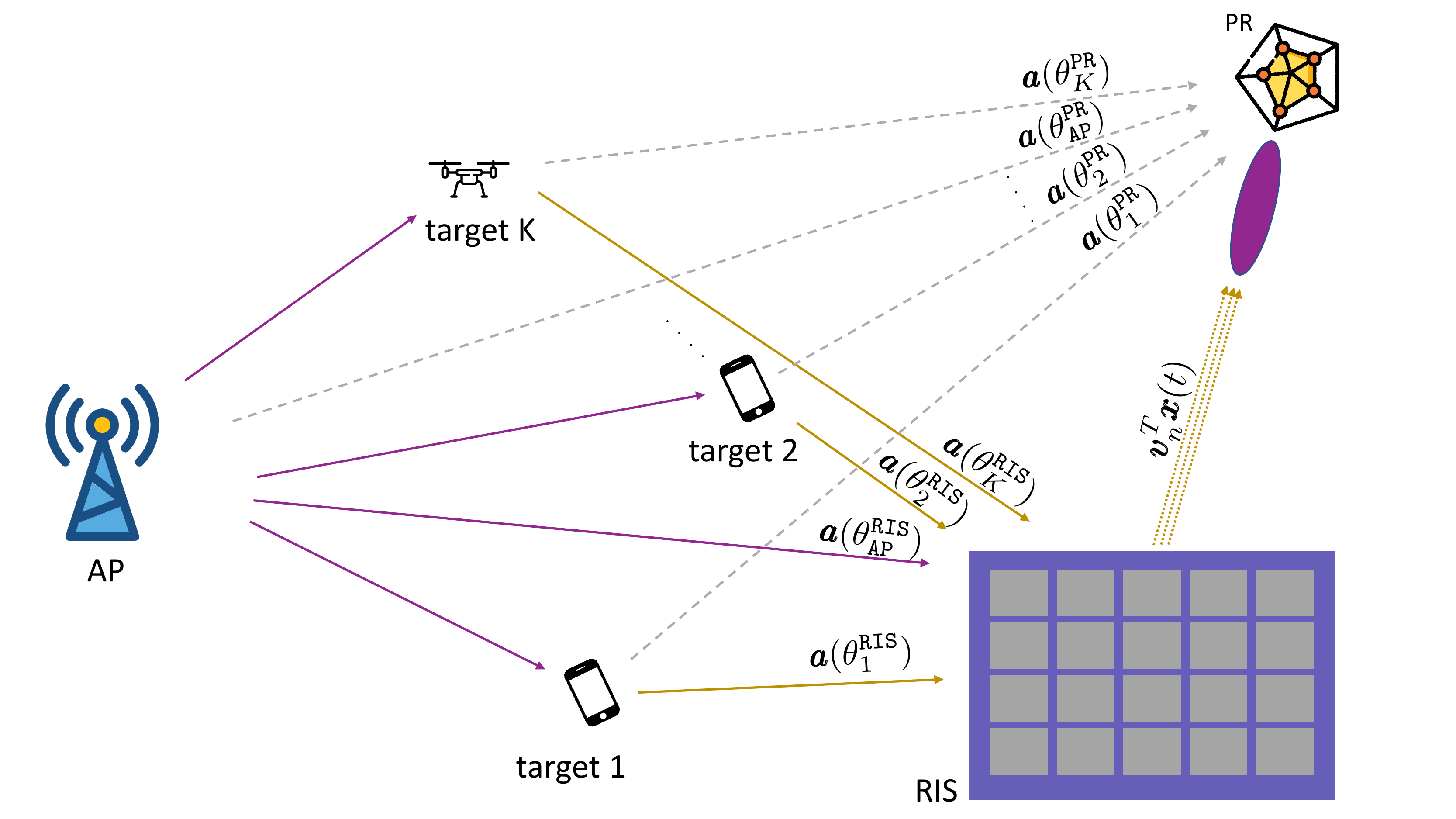}
\caption{The considered \ac{RIS}-based multi-target system comprising of an access point simultaneously serving communication users in the downlink, where the $M$-element \ac{RIS} reflects back the bounces off the targets towards the \ac{PR}. The \ac{RIS} is considered to be founded on the surrounding building’s facade and the \ac{PR} is in the vicinity. The transmit signal $s(t)$ propagates and bounces off the targets towards the \ac{PR}, enabled by the reconfigurable behavior of the \ac{RIS}. The signal $s(t)$ is any communication signal, which is also used as data to perform joint estimation and detection at the \ac{PR}. }
\label{fig_1}
\end{figure}
Let us assume a narrow-band system where $K$ targets are to be localized. Note that communication users will likewise be localized as they are also regarded as targets. For simplicity, assume that the base station is equipped with a single antenna and transmits a signal $s(t)$. Note that in a communication-centric \ac{ISAC} setting, the signal $s(t)$, which is meant for communications, is leveraged by the radar sub-systems as well, i.e. the \ac{PR}. In the downlink, since the signal $s(t)$, intended for a specific communication user, is transmitted by a base station via an omni-directional antenna, the signal $s(t)$ is broadcasted. The signal would then hit $K$ targets. The \ac{RIS} receives the bounces off the $K$ targets, as well as the direct path between the AP and itself. Therefore, the received signal at the \ac{RIS} reads
\begin{equation}
\label{eq:rt}
	\pmb{r}(t) = \pmb{A}(\Theta_{1:K}^{\tt{RIS}}) \pmb{s}(t)
	+
	\alpha_0
	 \pmb{a}_{M}(\theta_{\tt{AP}}^{\tt{RIS}})s(t - \tau_{\tt{AP}}^{\tt{RIS}}),
\end{equation}
where $\Theta_{1:K}^{\tt{RIS}} = \begin{bmatrix}
	\theta_1^{\tt{RIS}} & \theta_2^{\tt{RIS}} & \ldots & \theta_K^{\tt{RIS}}
\end{bmatrix}$ and $\theta_k^{\tt{RIS}}$ is the angle between the $k^{th}$ target and the \ac{RIS}. Furthermore, the angle $\theta_{\tt{AP}}^{\tt{RIS}}$ is the angle between the AP and the \ac{RIS}. The matrix $\pmb{A}(\Theta_{1:K}^{RIS})$ is the steering matrix, which results from the contribution of each path between the targets and the \ac{RIS}, namely
\begin{equation}
	\pmb{A}(\Theta_{1:K}^{\tt{RIS}}) 
	=
	\begin{bmatrix}
		\pmb{a}_{M}(\theta_{1}^{\tt{RIS}}) 
		& \pmb{a}_{M}(\theta_{2}^{\tt{RIS}}) 
		& 
		\ldots 
		&
		\pmb{a}_{M}(\theta_{K}^{\tt{RIS}})
	\end{bmatrix},
\end{equation} 
where $\pmb{a}_{M}(\theta) \in \mathbb{C}^{M \times 1}$ is the steering vector at the output of the $M$-elements of the \ac{RIS}. The concatenation of all steering vectors defines the array manifold, $\pmb{A}(\Theta_{1:K}^{\tt{RIS}}) \in \mathbb{C}^{M \times K}$. It is worth noting that this paper does not assume any structure on $\pmb{a}_{M}(\theta)$. The signal $\pmb{s}(t)$ contains scaled and delayed versions of the original transmitted signal $s(t)$, and is represented as 
\begin{equation}
	\pmb{s}(t)
	=
	\begin{bmatrix}
		\alpha_1 s(t - \tau_{\tt{AP}}^{1} -\tau_{1}^{\tt{RIS}}) \\
		\alpha_2 s(t - \tau_{\tt{AP}}^{2} -\tau_{2}^{\tt{RIS}}) \\
		\vdots \\
		\alpha_K s(t - \tau_{\tt{AP}}^{K} -\tau_{K}^{\tt{RIS}}) \\
	\end{bmatrix} \in \mathbb{C}^{K \times 1},
\end{equation}
where $\tau_{\tt{AP}}^{k}$ is the propagation delay between the AP and the $k^{th}$ target, $\tau_{k}^{\tt{RIS}}$ is the propagation delay between the $k^{th}$ target and the \ac{RIS}, and $\tau_{\tt{AP}}^{\tt{RIS}}$ is the propagation delay between the AP and the \ac{RIS}. Furthermore, the complex coefficient $\alpha_k$, for $k = 1 \ldots K$, is the reflection gain, taking into account large scale fading effects, resulting from the path between the AP towards the $k^{th}$ target then towards the \ac{RIS}. Likewise, $\alpha_0$ is the complex coefficient taking into account the channel gain between the AP and the \ac{RIS}.  Furthermore, the reflected signal from the \ac{RIS} towards the \ac{PR} is expressed as 
\begin{equation}
	x_n(t) 
	=
	\pmb{v}_n^T
	\Big(
	\pmb{A}(\Theta_{1:K}^{\tt{RIS}}) \pmb{s}(t)
	+
	\alpha_0
	 \pmb{a}_{M}(\theta_{\tt{AP}}^{\tt{RIS}})s(t - \tau_{\tt{AP}}^{\tt{RIS}})
	 \Big),
\end{equation}
where $\pmb{v}_n \in \mathbb{C}^{M \times 1}$ is the phase shifts due to the reflecting elements of the \ac{RIS} at the $n^{th}$ epoch. We define the epoch as a period of time receiving a realization of $\pmb{r}(t)$ defined in equation \eqref{eq:rt}.
The final received signal at the \ac{PR} is expressed over three counterparts as follows
\begin{equation}
\begin{split}
	\pmb{y}_n(t)
	&=
	\rho_{\tt{RIS}}^{\tt{PR}}\pmb{a}(\theta_{\tt{RIS}}^{\tt{PR}})x_n(t - \tau_{\tt{RIS}}^{\tt{PR}}) \\
	&+
	\rho_{\tt{AP}}^{\tt{PR}} \pmb{a}(\theta_{\tt{AP}}^{\tt{PR}})
	s(t-\tau_{\tt{AP}}^{\tt{PR}}) \\
	&+
	\sum\limits_{k=1}^K
	\rho_k
	\pmb{a}(\theta_{k}^{\tt{PR}})
	s(t-\tau_{\tt{AP}}^{k}-\tau_k^{\tt{PR}})+\pmb{\epsilon}_n(t),
\end{split}
\end{equation}
where $\pmb{y}_n(t) \in \mathbb{C}^{N_{PR} \times 1}$ is the received signal at the \ac{PR}. The first term, i.e. $\rho_{\tt{RIS}}^{\tt{PR}}\pmb{a}(\theta_{\tt{RIS}}^{\tt{PR}})x_n(t - \tau_{\tt{RIS}}^{\tt{PR}})$,  contains all information gathered at the \ac{RIS} and reflected back towards the \ac{PR}. The second term, namely $\rho_{\tt{AP}}^{\tt{PR}} \pmb{a}(\theta_{\tt{AP}}^{\tt{PR}})
	s(t-\tau_{\tt{AP}}^{\tt{PR}})$,  consists of the direct LoS between the AP and the PR. The third term, that is $\sum\limits_{k=1}^K
	\rho_k
	\pmb{a}(\theta_{k}^{\tt{PR}})
	s(t-\tau_{\tt{AP}}^{k}-\tau_k^{\tt{PR}})$, contains $K$ target contributions that propagated from the AP towards the targets then bounced off the targets towards the \ac{PR}. The propagation delay $\tau_{\tt{RIS}}^{\tt{PR}}$ is the delay between the \ac{RIS} and the \ac{PR}, $\tau_{\tt{AP}}^{\tt{PR}}$ is the delay between AP and the \ac{PR} and $\tau_k^{\tt{PR}}$ is the delay between the $k^{th}$ target and the \ac{PR}. Moreover, the large scale fading gains $\rho_k$'s are defined in a similar manner as $\alpha_k$'s. Furthermore, the noise term $\pmb{\epsilon}_n(t)$ is additive white Gaussian background noise. Sampling at $L$ time instances, we have 
\begin{equation}
\begin{split}
	\pmb{Y}_n
	&=
	\begin{bmatrix}
		\pmb{y}_n(1) & \pmb{y}_n(2) & \ldots & \pmb{y}_n(L)
	\end{bmatrix}  
	 \\
	&= 
	\rho_{\tt{RIS}}^{\tt{PR}}
	\pmb{a}(\theta_{\tt{RIS}}^{\tt{PR}})
	\pmb{x}_n
	+
	\rho_{\tt{AP}}^{\tt{PR}} 
\pmb{a}(\theta_{\tt{AP}}^{\tt{PR}})
	\pmb{s}_{\tt{AP}}^{\tt{PR}}
	+
\sum\limits_{k=1}^K
\rho_k
	\pmb{a}(\theta_{k}^{\tt{PR}})
	\pmb{s}_k^{\tt{PR}}
	+
	\pmb{E}_n,
\end{split}
\end{equation}
where $\pmb{Y}_n \in \mathbb{C}^{N_{\tt{PR}} \times L}$ is the sampled data matrix at the \ac{PR} at the $n^{th}$ epoch. Also,  
\begin{align}
	\pmb{x}_n &= 
	\begin{bmatrix}
		x_n(t_1) & x_n(t_2) & \ldots & x_n(t_L)
	\end{bmatrix}, \\
	\pmb{s}_{\tt{AP}}^{\tt{PR}} &= 
	\begin{bmatrix}
		s(t_1-\tau_{\tt{AP}}^{\tt{PR}}) \\
		s(t_2-\tau_{\tt{AP}}^{\tt{PR}}) \\
		\vdots \\
		s(t_L-\tau_{\tt{AP}}^{\tt{PR}})
	\end{bmatrix}^T ,\\
	\pmb{s}_k^{\tt{PR}} 
	&=
	\begin{bmatrix}
		s(t_1-\tau_{\tt{AP}}^{k}-\tau_k^{\tt{PR}})  \\
		s(t_2-\tau_{\tt{AP}}^{k}-\tau_k^{\tt{PR}})  \\
		\vdots \\
		s(t_L-\tau_{\tt{AP}}^{k}-\tau_k^{\tt{PR}}) 
	\end{bmatrix}^T ,\\
	\pmb{E}_n &= 
	\begin{bmatrix}
		\pmb{\epsilon}_n(t_1) & \pmb{\epsilon}_n(t_2) & \ldots & \pmb{\epsilon}_n(t_L)
	\end{bmatrix}^T .
\end{align}
Collecting all samples at different epochs in one data matrix, we now have the following at the \ac{PR},
\begin{equation}
	\pmb{Y} 
	=
	\begin{bmatrix}
		\pmb{Y}_1 \\
		\pmb{Y}_2 \\
		\vdots \\
		\pmb{Y}_N
	\end{bmatrix}
	\in
	\mathbb{C}^{N_{\text{PR}}N_{\text{epoch}} \times L}.
\end{equation}
Now, we are ready to address our problem, i.e. \textit{given our data matrix $\pmb{Y}$, which contains information over different trajectories, time instances and passive radar sensors, our objective is to estimate the angles of arrivals  of the targets that bounced off the \ac{RIS} towards the \ac{PR}.}

\section{Reflection Matrix Optimization}
\label{sec:reflection-matrix-optimization}
The RIS reflection matrix is used for radar-centric sensing performance enhancements. Our goal is to maximize the contributions of the $K$ targets collected at the RIS and minimize the power of all other contributions, which in this case come from the AP. At the $n^{th}$ epoch, one possible optimization problem would be to
\begin{equation}
 \label{eq:CancelAP-RIS}
\begin{aligned}
(\mathcal{P}_{\tt{RIS}}):
\begin{cases}
\min\limits_{ \pmb{v}_n }&  \Vert \pmb{v}_n^T \pmb{a}(\theta_{\tt{AP}}^{\tt{RIS}}) \Vert^2   \\
\textrm{s.t.}
 & \vert \pmb{v}_{n} \vert  = 1.
\end{cases}
\end{aligned}
\end{equation}
Note that in the above, we have decided not to create specific beams towards the \ac{PR}, as we can have multiple \ac{PR}s in the scene. Therefore, the objective aims at minimizing the known static contribution between AP and \ac{RIS}. Also, the above problem acts on an instantaneous basis, that is it optimizes the reflection coefficients of the RIS based on observations at the $n^{th}$ epoch only. To minimize over all epochs, we propose the following criterion,
\begin{equation}
 \label{eq:CancelAP-RIS-Batch}
\begin{aligned}
(\mathcal{P}_{\tt{RIS}}):
\begin{cases}
\min\limits_{ \pmb{V} }&  \Vert \pmb{V} \pmb{a}(\theta_{\tt{AP}}^{\tt{RIS}}) \Vert^2   \\
\textrm{s.t.}
 & \vert \pmb{V}_{i,j} \vert  = 1,
\end{cases}
\end{aligned}
\end{equation}
where $\pmb{V} = \begin{bmatrix} \pmb{v}_1  & \pmb{v}_2&   \ldots & \pmb{v}_{N_{\tt{epoch}}}\end{bmatrix}^T \in \mathbb{C}^{N_{\tt{epoch}} \times M}$. The above problem is obviously a non-convex optimization problem. To this end,  we propose to first relax the problem as follows: 1) solve the unconstrained optimization problem then 2) adapt the solution towards a feasible one. Following this approach, the solution of the unconstrained variant of $(\mathcal{P}_{\tt{RIS}})$ is 
\begin{equation}
	\overline{\pmb{V}} = \mathcal{P}^{\perp}_{\pmb{a}(\theta_{AP}^{RIS})} = \pmb{I}_M -\frac{1}{\Vert \pmb{a}(\theta_{AP}^{RIS}) \Vert^2 } \pmb{a}(\theta_{AP}^{RIS})\pmb{a}^H(\theta_{AP}^{RIS}),
\end{equation}
which constitutes a one-time computation. Next, the solution is in-feasible, since there is no guarantee that $\overline{\pmb{V}}_{i,j}$ satisfies the constant modulus constraint. 
\begin{equation}
	\pmb{V} = 
	\exp
	\Big\lbrace
	j \angle (\mathcal{P}^{\perp}_{\pmb{a}(\theta_{AP}^{RIS})} \pmb{\Gamma} )
	\Big\rbrace,
\end{equation}
where $\pmb{\Gamma} \in \mathbb{C}^{M \times N_{\text{epoch}}}$ is a random matrix drawn from a standard Gaussian distribution. 
Note that the solution, $\pmb{V}$ is independent of the signal $s(t)$, hence no signal knowledge is required at the \ac{RIS}, as well as the \ac{PR}.


\section{Beamforming towards RIS}
\label{sec:beamforming-towards-RIS}
First, let us assume that the \ac{PR} beamforms towards the \ac{RIS} using the same beamforming vector $\pmb{w}$ over all time epochs as follows
\begin{equation}
\label{eq:Zeq}
	\pmb{Z} 
	=
	\begin{bmatrix}
		\pmb{w}^H\pmb{Y}_1 \\
		\pmb{w}^H\pmb{Y}_2 \\
		\vdots \\
		\pmb{w}^H\pmb{Y}_N
	\end{bmatrix}
	=
	\begin{bmatrix}
		\pmb{z}_1 & \ldots & \pmb{z}_L
	\end{bmatrix}
	\in 
	\mathbb{C}^{N_{\tt{epoch}} \times L}.
\end{equation}
Notice that the output of the beamformer at any epoch $n$ is given as follows
\begin{equation}
\begin{split}
	\pmb{w}^H\pmb{Y}_n
	&=
	\rho_{\tt{RIS}}^{\tt{PR}}
\pmb{w}^H
	\pmb{a}(\theta_{\tt{RIS}}^{\tt{PR}})
	\pmb{x}_n \\
	&+
\pmb{w}^H
\underbrace{
\Big(
	\rho_{\tt{AP}}^{\tt{PR}} 
\pmb{a}(\theta_{\tt{AP}}^{\tt{PR}})
	\pmb{s}_{\tt{AP}}^{\tt{PR}}
	+
\sum\limits_{k=1}^K
\rho_k
	\pmb{a}(\theta_{k}^{\tt{PR}})
	\pmb{s}_k^{\tt{PR}}
	+
	\pmb{E}_n
	\Big)
	}_{\text{interference + noise}}.
\end{split}
\end{equation}
Aiming at maximizing the output power over all epochs, we can propose the following optimization problem
\begin{equation}
 \label{eq:MVDR}
\begin{aligned}
(\mathcal{P}_1):
\begin{cases}
\max\limits_{ \pmb{w} }&   \sum\limits_{n=1}^N \pmb{R}_{x_nx_n}  \\
\textrm{s.t.}
 &  \pmb{w}^H\pmb{a}(\theta_{\tt{RIS}}^{\tt{PR}}) = 1,
\end{cases}
\end{aligned}
\end{equation}
where $\pmb{R}_{x_nx_n}$ is the output power contribution of the useful part looking in the direction of $\theta_{\tt{RIS}}^{\tt{PR}}$ 
\begin{equation}
	\pmb{R}_{x_nx_n}
	=
	\vert \alpha_n \vert^2
\pmb{w}^H\pmb{a}(\theta_{\tt{RIS}}^{\tt{PR}})
	\pmb{a}^H(\theta_{\tt{RIS}}^{\tt{PR}})	
\pmb{w},
\end{equation}
where $\alpha_n = \rho_{\tt{RIS}}^{\tt{PR}}\sqrt{	\pmb{x}_n^T\pmb{x}_n^*}$.  The Lagrangian function associated with the above optimization problem is given as follows
\begin{equation}
	\mathcal{L}(\pmb{w},\lambda)
	=
	\vert \alpha \vert^2 
\pmb{w}^H\pmb{a}(\theta_{\tt{RIS}}^{\tt{PR}})
	\pmb{a}^H(\theta_{\tt{RIS}}^{\tt{PR}})	
\pmb{w}
-
\lambda 
(\pmb{w}^H\pmb{a}(\theta_{\tt{RIS}}^{\tt{PR}}) - 1),
\end{equation}
where $\vert \alpha \vert^2  = \sum\limits_{n=1}^N \vert \alpha_n \vert^2 $. Deriving with respect to $\pmb{w}$, we get the following expression
\begin{equation}
	\frac{\partial}{\partial \pmb{w}}
	\mathcal{L}(\pmb{w},\lambda)
	=
2\vert \alpha \vert^2 \pmb{a}(\theta_{\tt{RIS}}^{\tt{PR}})
	\pmb{a}^H(\theta_{\tt{RIS}}^{\tt{PR}})	
\pmb{w}
-
\lambda
\pmb{a}(\theta_{\tt{RIS}}^{\tt{PR}})	,
\end{equation}
which is null at 
\begin{equation}
\label{eq:beamformweight}
	\pmb{w}
	=
	\frac{\pmb{a}(\theta_{\tt{RIS}}^{\tt{PR}})}{\Vert \pmb{a}(\theta_{\tt{RIS}}^{\tt{PR}}) \Vert^2 },
\end{equation}
and 
\begin{equation}
	\lambda
	=
	2\vert \alpha \vert^2.
\end{equation}
To this end, and after beamforming via equation \eqref{eq:beamformweight}, our model becomes
\begin{equation}
	\pmb{Z}_n 
	= \pmb{w}^H \pmb{Y}_n 
	=\rho_{\tt{RIS}}^{\tt{PR}} \pmb{x}_n + \pmb{i}_n + \tilde{\pmb{\epsilon}}_n,
\end{equation}
with
\begin{equation}
	\pmb{i}_n
	=
	\rho_{\tt{AP}}^{\tt{PR}}\beta(\theta_{\tt{AP}}^{\tt{PR}})\pmb{s}_{\tt{AP}}^{\tt{PR}} 
	+
	\sum\limits_{k=1}^K
	\rho_k
	\beta(\theta_k^{\tt{PR}})
	\pmb{s}_k^{\tt{PR}},
\end{equation}
where $\tilde{\pmb{\epsilon}}_n = \pmb{w}^H \pmb{E}_n$ and $\beta(\theta) = \pmb{w}^H \pmb{a}(\theta)$ is the output of the beampattern used with our beamforming vector to maximize the power in the direction of $\theta_{\tt{RIS}}^{\tt{PR}}$. With increasing number of antennas for ULAs, it is easily verified that $\beta(\theta) \xrightarrow[N_{PR} \rightarrow \infty]{} 0$ for all $\theta \neq \theta_{\tt{RIS}}^{\tt{PR}}$. Therefore, this means that all signal contributions that are not reflected by the \ac{RIS}, $\pmb{i}_n \xrightarrow[N_{PR} \rightarrow \infty]{} \pmb{0}$. After beamforming at each time epoch, the collected data are stacked as
\begin{equation}
\begin{split}
	\pmb{Z}
	=
	\begin{bmatrix}
		\pmb{w}^H \pmb{Y}_1  \\
		\pmb{w}^H \pmb{Y}_2  \\
		\vdots \\
		\pmb{w}^H \pmb{Y}_{N_{\tt{epoch}}}
	\end{bmatrix}
	=
	\begin{bmatrix}
		\pmb{z}_1 & \pmb{z}_2 & \ldots & \pmb{z}_L
	\end{bmatrix},
\end{split}
\end{equation}
where $\pmb{Z} \in \mathbb{C}^{N_{\tt{epoch}} \times L}$ is the beamformed data matrix at the \ac{PR} over all epochs and time samples. In the next section, we describe two methods that process $\pmb{Z}$ in two different manners, i.e. as a batch or sequentially.

\section{Target Localization}
\label{sec:target-localization}
Given a matrix $\pmb{Z}$ at the passive radar, the task is to estimate all angles within $\pmb{\Theta}_{1:K}^{\tt{RIS}}$, which consists of all target \ac{AoA} information. However, in many situations, such as dynamic environments, the number of targets, i.e. $K$, may be unknown. To this end, consider the output of the sum-and-delay beamformer at direction $\theta$ utilizing the steering vector $\pmb{V}\pmb{a}(\theta)$ at the $\ell^{th}$ snapshot, namely
\begin{equation}
	p_{\ell}(\theta) 
	=
	\pmb{a}^H(\theta)
	\pmb{V}^H
	\pmb{z}_\ell.
\end{equation}
Then, we can define the error cost function that we try to minimize at each snapshot, i.e.
\begin{equation}
	C(\ell) = \mathbb{E}( \vert e(\ell) \vert^2 ),
\end{equation}
where $e(\ell) = p_{\ell}(\theta) - \pmb{a}^H \pmb{z}_\ell$. The least-mean square aims at minimizing the above cost. First, we take the gradient with respect to $\pmb{a}$ to get 
\begin{equation}
	\begin{split}
		\nabla_{\pmb{a}} C(\ell) &= 2 \mathbb{E} \lbrace \nabla_{\pmb{a}} [ e(\ell) ]e^*(\ell) \rbrace \\
								 &= -2 \mathbb{E} \lbrace ( p_{\ell}(\theta) - \pmb{a}^H \pmb{z}_\ell )^* \pmb{z}_\ell \rbrace,
	\end{split}
\end{equation}
where the last step is due to the gradient's value, $\nabla_{\pmb{a}} [ e(\ell) ] = -\pmb{z}_\ell$. Thanks to $\nabla_{\pmb{a}} C(\ell) $, the \ac{LMS} filter can now focus towards the steepest ascent of $C(\ell)$ by taking a direction opposite to $\nabla_{\pmb{a}} C(\ell)$, namely
\begin{equation}
\begin{split}
	\hat{\pmb{a}}_{\ell + 1}(\theta)
	&=
	\hat{\pmb{a}}_{\ell}(\theta)
	-
	\frac{\mu}{2} \big[ \nabla_{\pmb{a}} C(\ell) \big]_{\pmb{a} = \hat{\pmb{a}}_{\ell}} \\
	&= \hat{\pmb{a}}_{\ell}(\theta)
	+
	\mu \mathbb{E} \lbrace ( p_{\ell}(\theta) - \hat{\pmb{a}}_{\ell}^H(\theta) \pmb{z}_\ell )^* \pmb{z}_\ell \rbrace,
\end{split}
\end{equation}
where $\frac{\mu}{2}$ is the well-known \ac{LMS} adaptation constant, i.e. the step size. Now that we have an update equation for the steering vector in the $\theta$'s look-direction, the only problem that remains is the expectation operator, which is not available at time update $\ell$. Therefore, omitting it we finally have
\begin{equation}
	\hat{\pmb{a}}_{\ell + 1}(\theta)
	=
	\hat{\pmb{a}}_{\ell}(\theta)
	+
	\mu ( p_{\ell}(\theta) - \hat{\pmb{a}}_{\ell}^H(\theta) \pmb{z}_\ell )^* \pmb{z}_\ell .
\end{equation}
A normalized \ac{LMS}, i.e. \ac{NLMS}, could also be proposed at this point to avoid sensitivities of the input norm, i.e. $\Vert \pmb{z}_\ell \Vert$, which makes it extremely difficult to choose an adaptation constant for algorithm stability purposes \cite{haykin1996adaptive}. To this end, the following update is proposed
\begin{equation}
	\hat{\pmb{a}}_{\ell + 1}(\theta)
	=
	\hat{\pmb{a}}_{\ell}(\theta)
	+
	\frac{\mu}{\Vert \pmb{z}_{\ell} \Vert} ( p_{\ell}(\theta) - \hat{\pmb{a}}_{\ell}^H(\theta) \pmb{z}_\ell )^* \pmb{z}_\ell .
\end{equation}
Since $\hat{\pmb{a}}(\theta)$ also contains signal contributions present on direction $\theta$, then a reasonable spectrum would be to measure the total power contained in the final estimate $\hat{\pmb{a}}_{L}(\theta)$, namely
\begin{equation}
	P(\theta) = \Vert \hat{\pmb{a}}_{L}(\theta) \Vert^2 .
\end{equation}
Finally, the \ac{AoAs} corresponding to the $K$ targets are computed through a one-dimensional peak-finding search of $P(\theta)$, viz.
\begin{equation}
	\hat{\pmb{\Theta}}_{1:K}^{\tt{RIS}}
	=
	\arg\max_{\theta_1 \ldots \theta_K} P(\theta).
\end{equation}
A summary of an implementation of the proposed \ac{LMS} filtering approach is summarized in {\tt{\textbf{Algorithm \ref{alg:cap}}}}. One interesting feature of this batch approach is that it could be parallelized over angles to be evaluated on a grid, i.e. $\Theta_{\tt{grid}}$.

\begin{algorithm}
\caption{\tt{Batch NLMS filter Localization}}\label{alg:cap}
\begin{algorithmic}
\State \tt{\textbf{INPUT}} $\pmb{Z} = \begin{bmatrix} \pmb{z}_1 & \ldots & \pmb{z}_L \end{bmatrix}$, $\mu$, $\Theta_{\tt{grid}}$
\While{$\theta \neq \emptyset$}
\State $\hat{\pmb{a}}_{0}(\theta) \gets \pmb{0}_{N_{\tt{epoch}} \times 1}$  
\State $\ell \gets 0$
\While{$\ell < L$}
\State 	$p_{\ell}(\theta) 
	\gets
	\pmb{a}^H(\theta)
	\pmb{V}^H
	\pmb{z}_\ell$
\State $  \hat{\pmb{a}}_{\ell + 1}(\theta)
	\gets
	\hat{\pmb{a}}_{\ell}(\theta)
	+
	\frac{\mu}{\Vert \pmb{z}_{\ell} \Vert} ( p_{\ell}(\theta) - \hat{\pmb{a}}_{\ell}^H(\theta) \pmb{z}_\ell )^* \pmb{z}_\ell $
\State $\ell \gets \ell + 1$
\EndWhile
\State$P(\theta) = \Vert \hat{\pmb{a}}_{L}(\theta) \Vert^2 $
\State Choose next $\theta \in \Theta_{\tt{grid}}$
\EndWhile
\If{$P(\theta)$ \tt{is peak}} 
\State \tt{Insert} $\theta$ in $\hat{\pmb{\Theta}}_{1:K}^{\tt{RIS}}$
\EndIf
\State \\
\Return $\hat{\pmb{\Theta}}_{1:K}^{\tt{RIS}}$
\end{algorithmic}
\end{algorithm}
Note that for peak detection, we first normalize the entire spectrum $P(\theta)$ so that its maximum value is $1$. Then, we use MATLAB's $\tt{findpeaks()}$ function to determine the peaks that exceed a certain threshold value, say $ 0 < \varphi < 1$.

Even though the implementation of {\tt{\textbf{Algorithm \ref{alg:cap}}}}  is sequential over snapshot dimensions, i.e. over the columns of $\pmb{Z}$, one would have to wait for the entire samples to be received over epochs, i.e. the entire snapshot $\pmb{z}_\ell$ is needed so that the LMS algorithm can operate. In this context, a purely sequential approximated \ac{NLMS} is proposed. The advantage of such an implementation is evident, i.e. an estimate of the different \ac{AoAs} can be achieved per epoch. 

\begin{algorithm}
\caption{\tt{Sequential NLMS filter}}\label{alg:cap2}
\begin{algorithmic}
\State \tt{\textbf{INPUT}} $\pmb{Z} = \begin{bmatrix} \pmb{z}_1 & \ldots & \pmb{z}_L \end{bmatrix}$, $\mu$, $\Theta_{\tt{grid}}$
\State \tt{\textbf{INIT}} $d_{\ell}(\theta) \gets \pmb{0}_{N_{\tt{grid}} \times L}$, $P(\theta) \gets \pmb{0}_{N_{\tt{grid}} \times 1}$
\While{$n = 1 \ldots N_{\tt{epoch}}$} 
\State $\hat{\pmb{\Theta}}_{1:K,(n)}^{\tt{RIS}} = \emptyset$
\While{$\theta \neq \emptyset$}
\State $\ell \gets 0$
\State $p \gets 0$
\While{$\ell < L$}
\State  $d_{\ell}(\theta) \leftarrow d_{\ell}(\theta) + \pmb{a}^H(\theta) \pmb{V}_{n,:}^H \pmb{Z}_{n,\ell}  $
\State  $p \leftarrow p + \frac{\mu}{\Vert \pmb{Z}_{1:n,\ell} \Vert^2}(d_{\ell}(\theta) - p^* \pmb{Z}_{n,\ell} )^* \pmb{Z}_{n,\ell} $

\State $\ell \gets \ell + 1$
\EndWhile
\State$P(\theta) \leftarrow P(\theta) +  \vert p \vert^2 $
\State Choose next $\theta \in \Theta_{\tt{grid}}$
\EndWhile
\If{$P(\theta)$ \tt{is peak}} 
\State \tt{Insert} $\theta$ in $\hat{\pmb{\Theta}}_{1:K}^{{\tt{RIS}},(n)}$
\EndIf
\EndWhile
\State \\
\Return $\hat{\pmb{\Theta}}_{1:K}^{{\tt{RIS}},(n)}$
\end{algorithmic}
\end{algorithm}
It is worth noting that for each epoch $n$, {\tt{\textbf{Algorithm \ref{alg:cap2}}}} provides an estimate of the \ac{AoAs}, i.e. $\hat{\pmb{\Theta}}_{1:K}^{{\tt{RIS}},(n)}$, as opposed to {\tt{\textbf{Algorithm \ref{alg:cap}}}}, where we have to wait until the last epoch.

\section{Simulation Results}
\label{sec:simulations}
In this section, we present different simulation results, conducted to illustrate the performance of the proposed methods. Furthermore, these methods are compared with a baseline, running the same methods depicted in {\tt{\textbf{Algorithm \ref{alg:cap}}}} and {\tt{\textbf{Algorithm \ref{alg:cap2}}}}, but without an \ac{RIS}. Unless otherwise stated, we fix the number of epochs used for \ac{RIS} measurements to $N_{\tt{epoch}} = 100$. The number of time samples is set to $L= 100$ samples. Furthermore, the number of antenna elements at the \ac{PR} is $N_{\tt{PR}} = 8$ antennas. The threshold used for peak detection for both batch and sequential \ac{NLMS} methods described in {\tt{\textbf{Algorithm \ref{alg:cap}}}} and {\tt{\textbf{Algorithm \ref{alg:cap2}}}} is set to $\varphi = 0.5$. Moreover, we fix the angles between the AP and the \ac{RIS},  and the \ac{RIS} and the \ac{PR} to $\theta_{\tt{AP}}^{\tt{RIS}} = -10^\circ$ and $\theta_{\tt{RIS}}^{\tt{PR}} = -40^\circ$, respectively. A \ac{ULA} configuration spaced at half a wavelength has been integrated at the \ac{PR}.

\begin{figure}[t]
	\centering
	\includegraphics[width=1\linewidth]{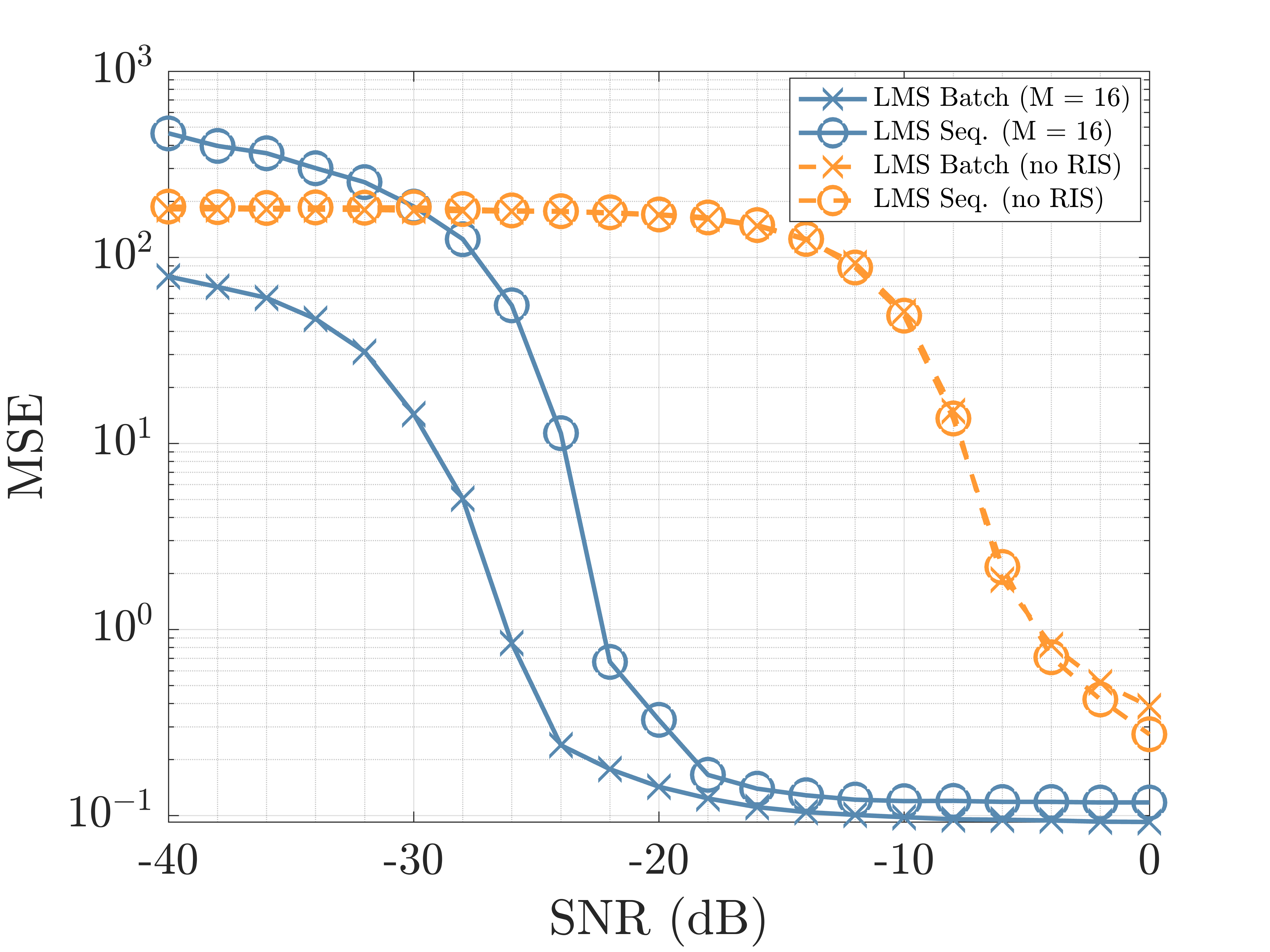}
	\caption{The \ac{MSE} vs. \ac{SNR} performances with $K = 2$ targets, for a \ac{ULA} configuration of $N_{\tt{PR}} = 8$ antennas and $L = 100$ time samples.}
	\label{fig:MSEvsSNR}
\end{figure}

\begin{figure}[t]
	\centering
	\includegraphics[width=1\linewidth]{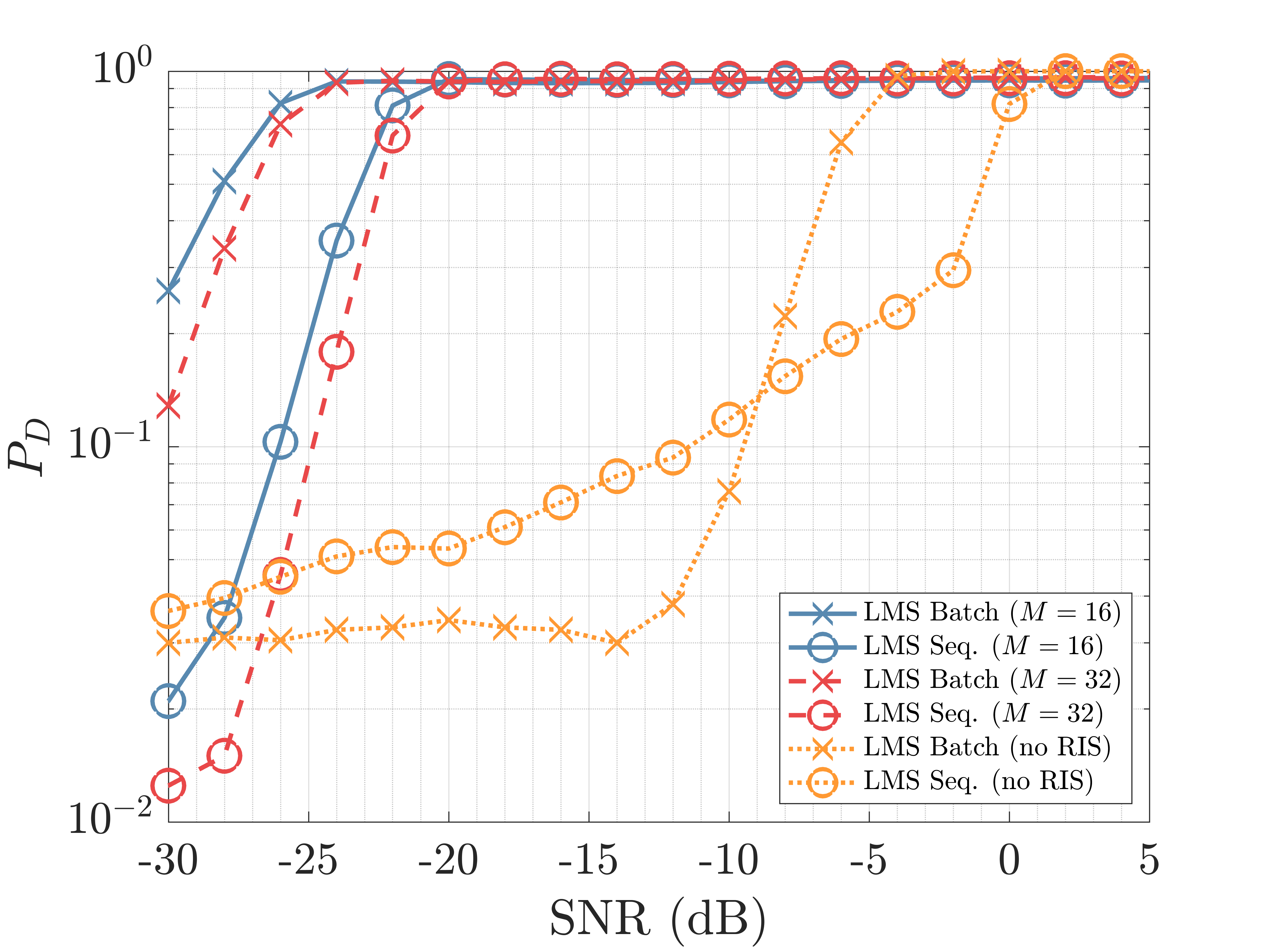}
	\caption{The probability of detection vs. \ac{SNR} performances with $N_{\tt{PR}} = 8$ antennas and $L = 100$ time samples. Two targets were set in the scene.}
	\label{fig:PDvsSNR}
\end{figure}

\begin{figure}[t]
	\centering
	\includegraphics[width=1\linewidth]{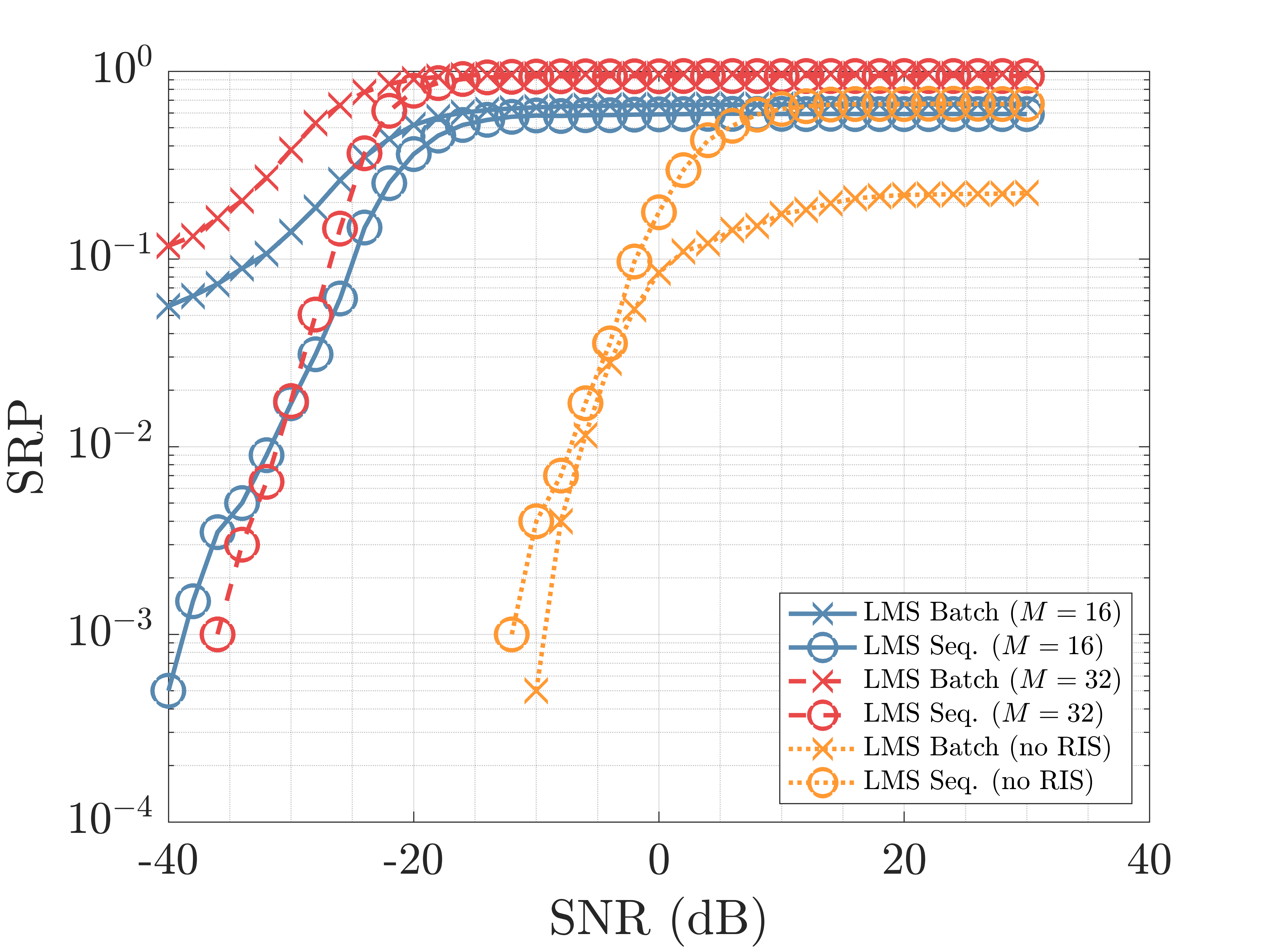}
	\caption{The \ac{SRP} vs. \ac{SNR} performances with $L = 100$ time samples, two targets and $N_{\tt{PR}} = 8$ antennas.}
	\label{fig:SRPvsSNR}
\end{figure}

\begin{figure}[t]
	\centering
	\includegraphics[width=1\linewidth]{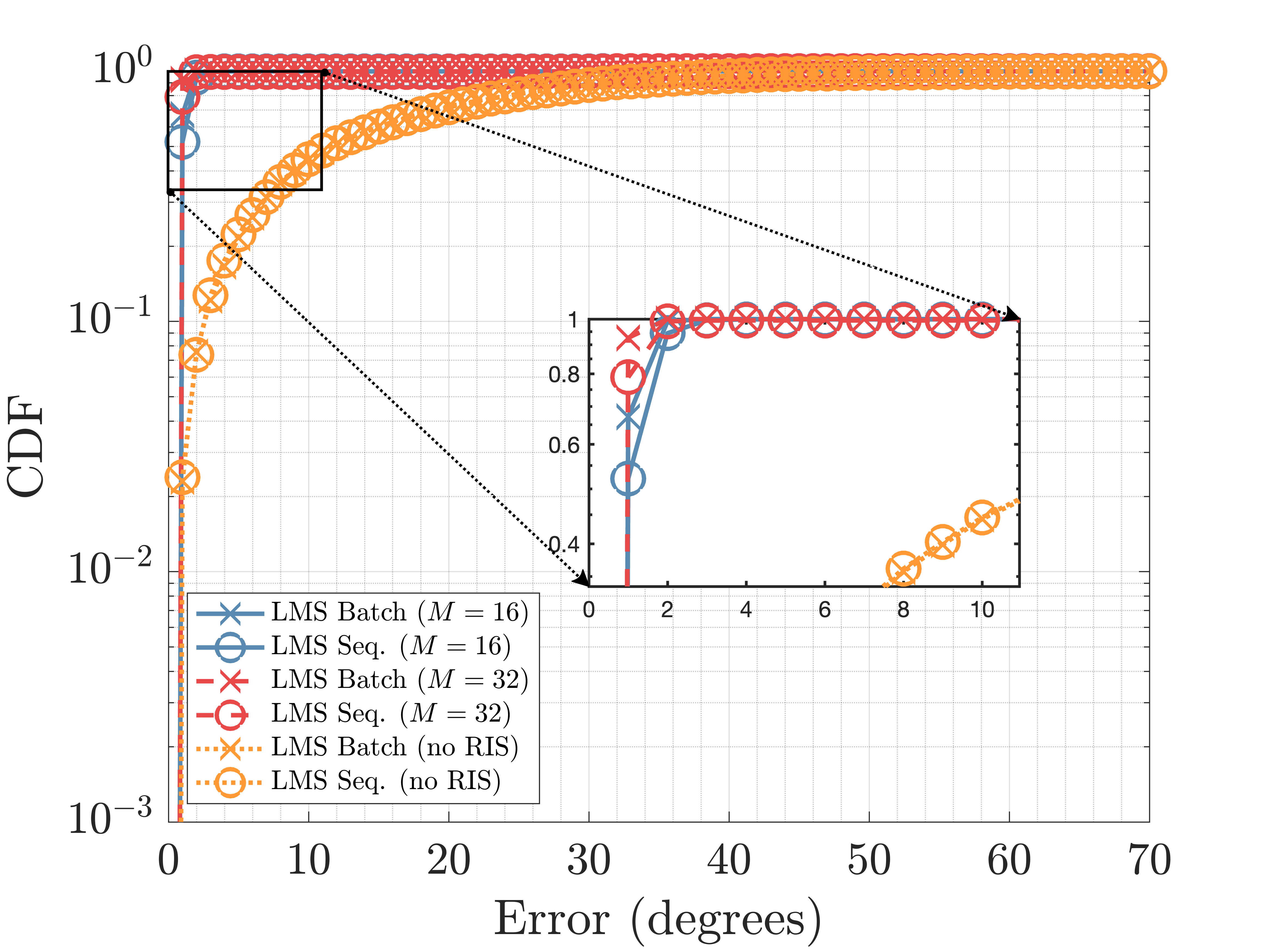}
	\caption{The \ac{CDF} performance with $K = 2$ targets, for a \ac{ULA} configuration of $N_{\tt{PR}} = 8$ antennas and $L = 100$ time samples.}
	\label{fig:CDFvsError}
\end{figure}

\begin{figure}[t]
	\centering
	\includegraphics[width=1\linewidth]{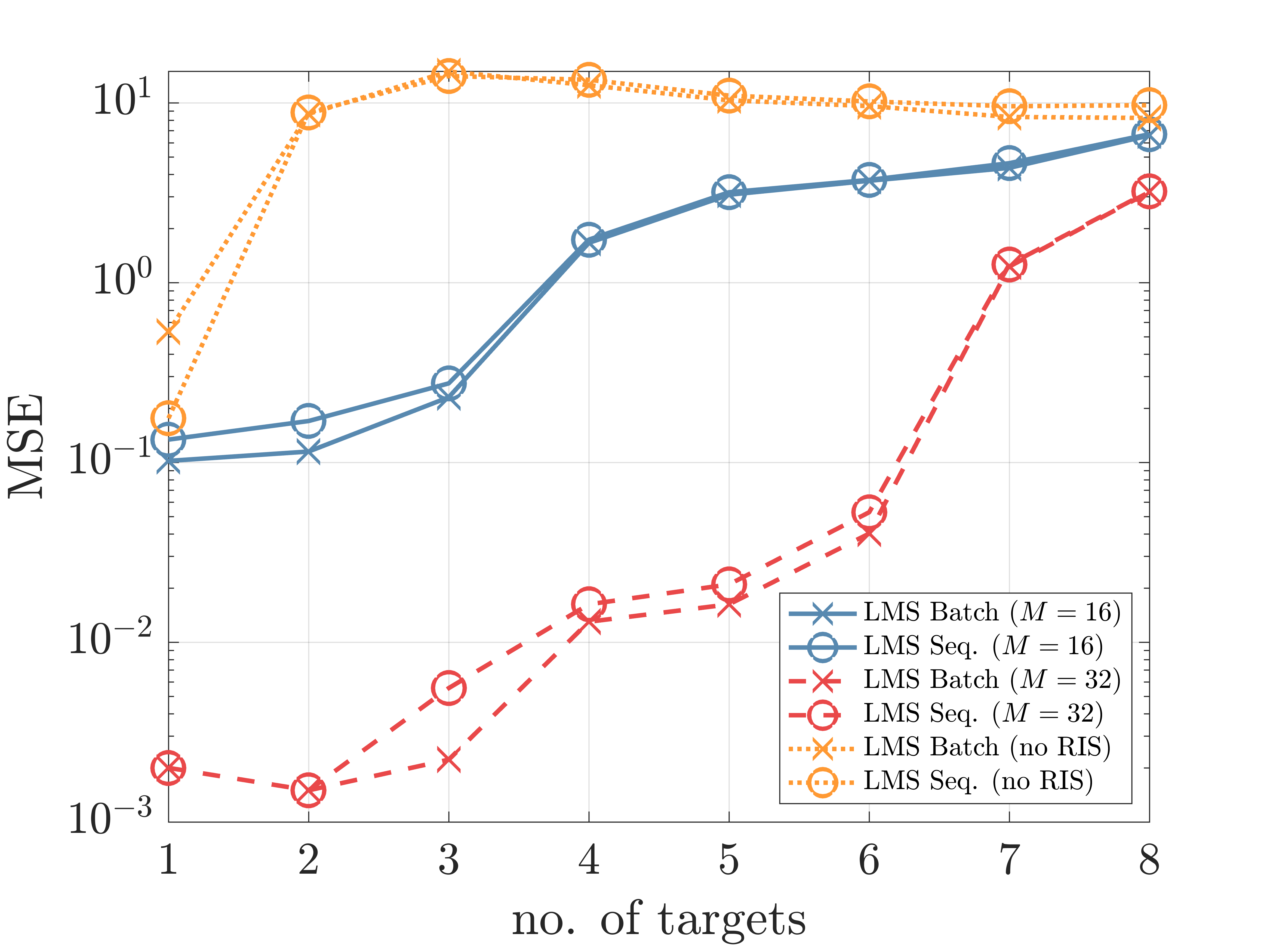}
	\caption{The \ac{MSE} behavior against the number of targets $K$ with fixed $\SNR = 20\dB$ at the \ac{PR}. }
	\label{fig:MSEvsNoOfTargets}
\end{figure}

\begin{figure}[t]
	\centering
	\includegraphics[width=1\linewidth]{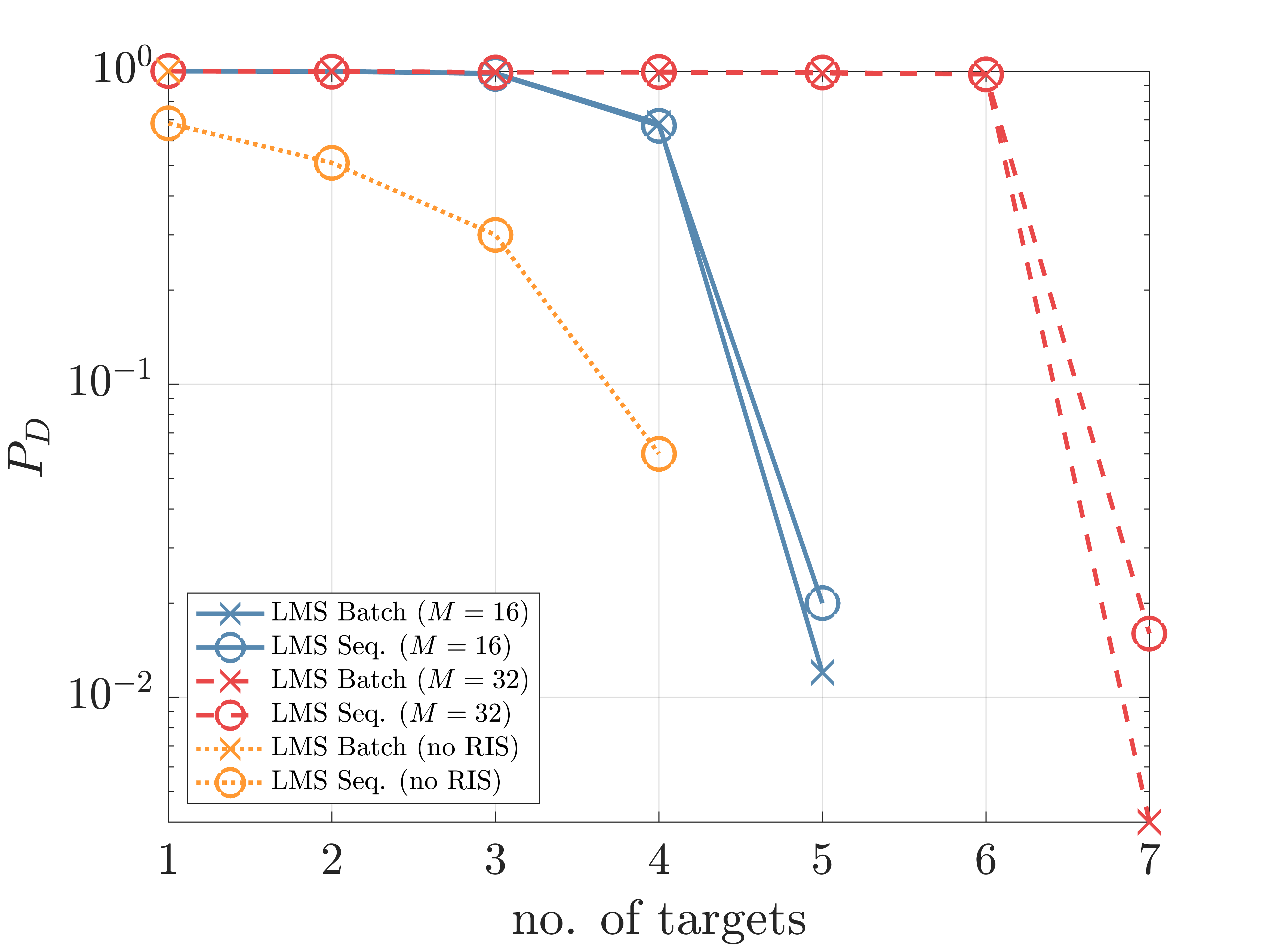}
	\caption{The detection probability behavior vs. $K$ with fixed $\SNR = 20\dB$ at the \ac{PR}.}
	\label{fig:PDvsNoOfTargets}
\end{figure}

\begin{figure}[t]
	\centering
	\includegraphics[width=1\linewidth]{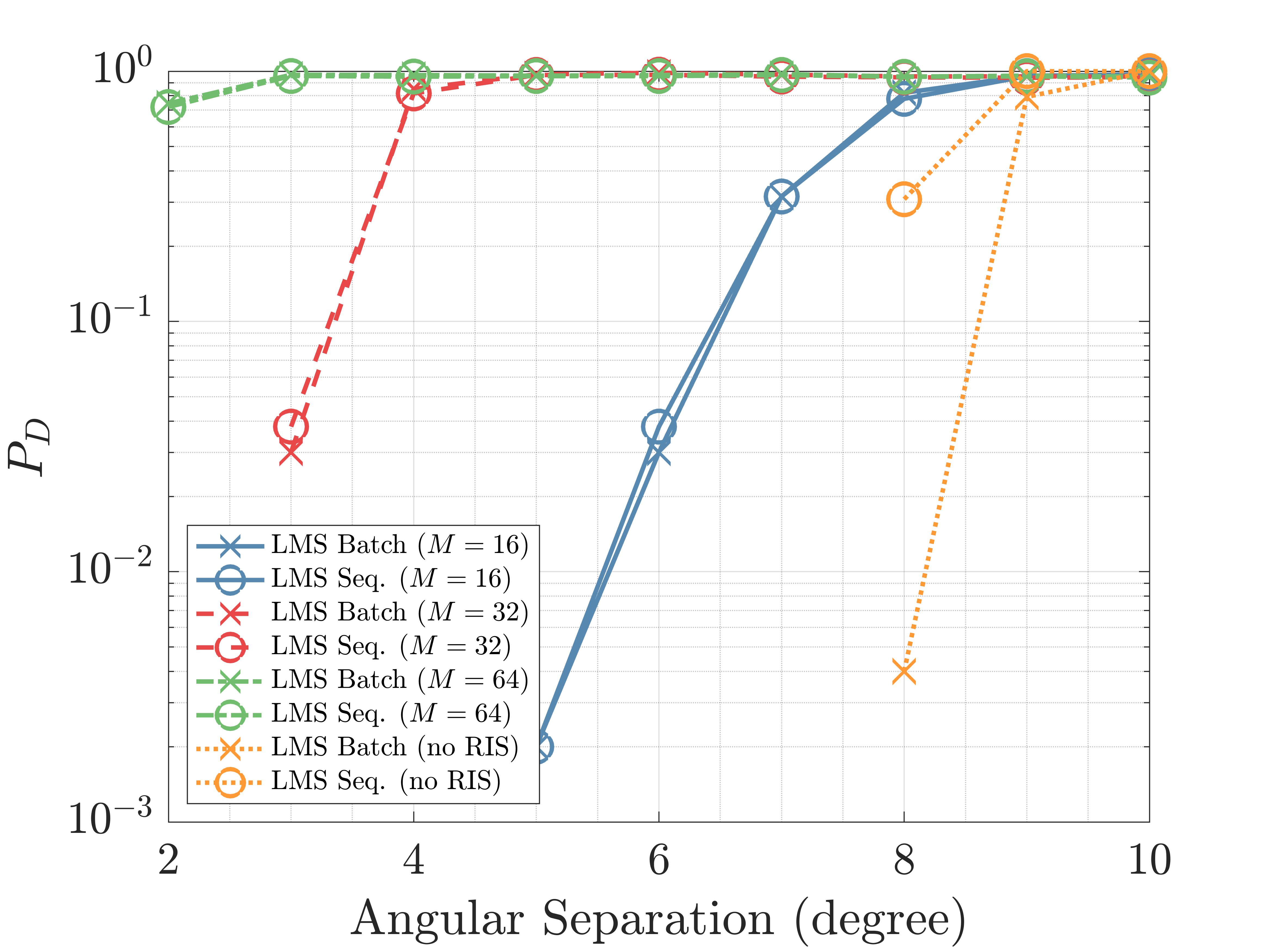}
	\caption{Detection probability vs. angular separation }
	\label{fig:PDvsAngularSeparation}
\end{figure}

\begin{figure*}[!t]
\centering
\subfloat[{\tt{\textbf{Algorithm \ref{alg:cap}}}}: \ac{NLMS} batch method]{\includegraphics[height=2.25in,width=3.5in]{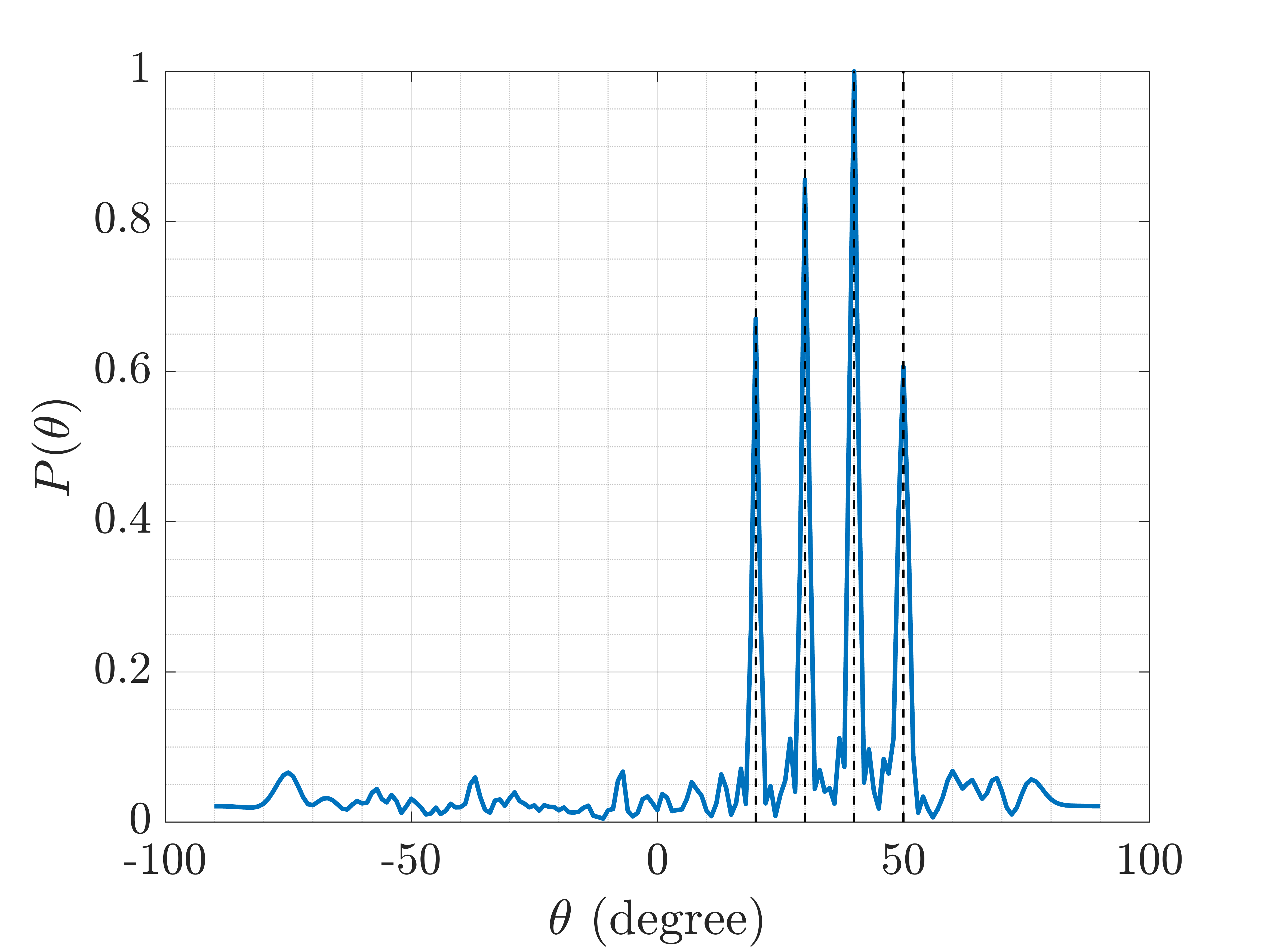} 
\label{fig:spectrum-LMS-batch}}
\hfil
\subfloat[{\tt{\textbf{Algorithm \ref{alg:cap2}}}}: \ac{NLMS} sequential method]{\includegraphics[height=2.25in,width=3.5in]{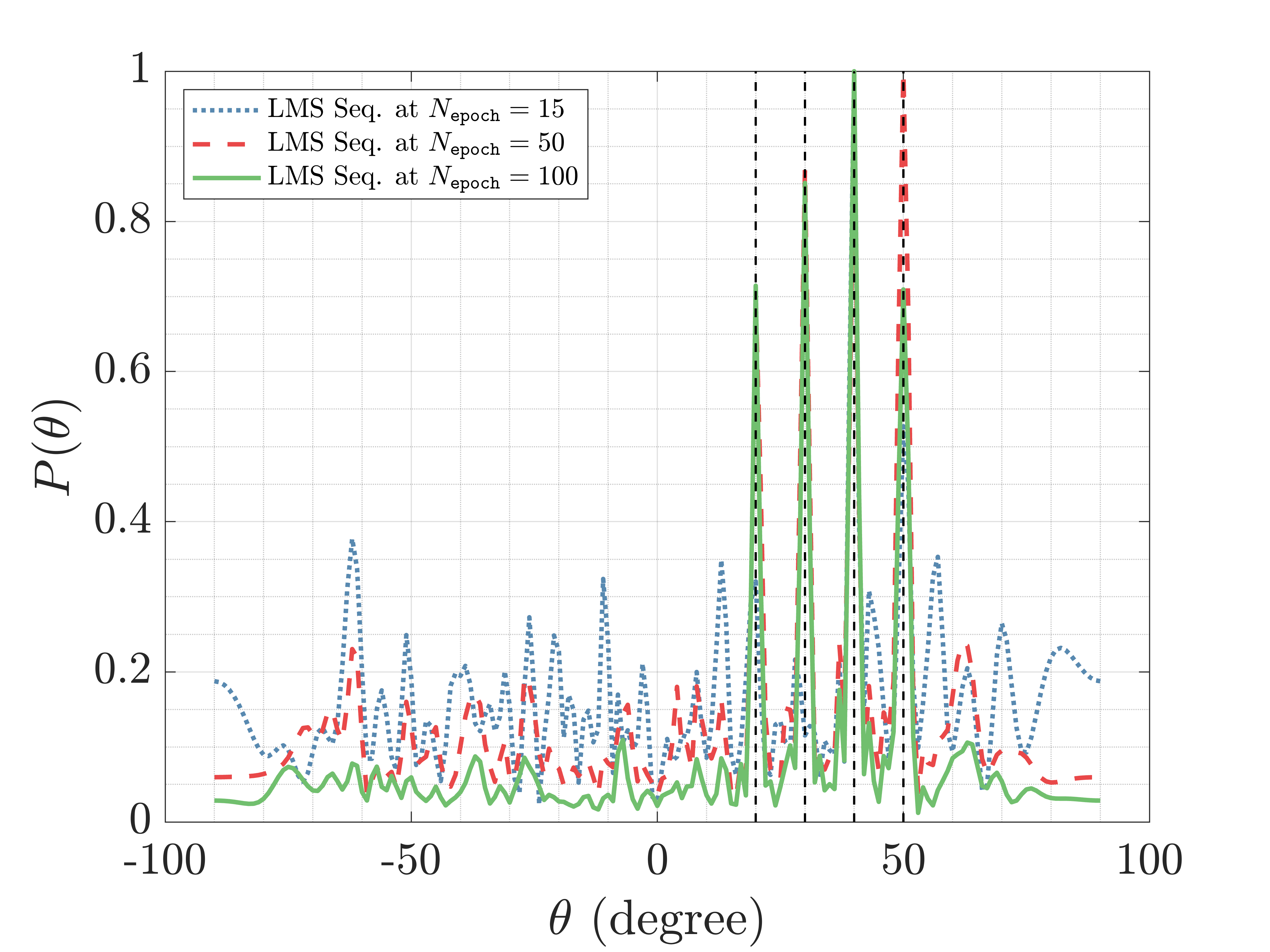}  
\label{fig:spectrum-LMS-seq}}
\caption{The spectra of the proposed \ac{NLMS} filtering methods utilizing the \ac{RIS}-aided architecture. The scene consists of $K = 4$ targets located at $\theta_k^{\tt{RIS}} = 20 + 10(k-1), \forall k$. The \ac{RIS} is equipped with $M = 64$ reflective elements and the number of time samples is $L = 100$. The number of epochs is set to $N_{\tt{epoch}} = 100$. Dashed vertical lines represent the true \ac{AoAs}.}
\label{fig:spectra}
\end{figure*}


In Fig. \ref{fig:MSEvsSNR}, we aim at analyzing the \ac{MSE} as a function of \ac{SNR} at the \ac{PR}. The \ac{MSE} is computed as 
\begin{equation}
	\MSE = \frac{1}{PK} \sum\limits_{k=1}^K \sum\limits_{p = 1}^{P} (\theta_k - \hat{\theta}_k)^2,
\end{equation}
The \ac{SNR} at the \ac{PR} is defined as 
\begin{equation}
\SNR = \frac{\mathbb{E} \big(\big\vert \pmb{y}^{(\tt{RIS})}_n(t) +   \pmb{y}^{(\tt{AP})}(t) +  \sum\limits_{k=1}^K \pmb{y}^{(k)}(t) \big\vert^2 \big)}{\mathbb{E}\big(\big\vert \pmb{\epsilon}_n(t) \big\vert^2 \big)},
\end{equation}
where
\begin{align}
	\pmb{y}^{(\tt{RIS})}_n(t)  &= \rho_{\tt{RIS}}^{\tt{PR}}\pmb{a}(\theta_{\tt{RIS}}^{\tt{PR}})x_n(t - \tau_{\tt{RIS}}^{\tt{PR}}), \\
	\pmb{y}^{(\tt{AP})}(t) &= \rho_{\tt{AP}}^{\tt{PR}} \pmb{a}(\theta_{\tt{AP}}^{\tt{PR}})s(t-\tau_{\tt{AP}}^{\tt{PR}}), \\
	\pmb{y}^{(k)}(t) &= \rho_k \pmb{a}(\theta_{k}^{\tt{PR}})s(t-\tau_{\tt{AP}}^{k}-\tau_k^{\tt{PR}}).
\end{align}
The parameter $\theta_k$ is the true \ac{AoA} of the $k^{th}$ target and $\hat{\theta}_k$ is the estimated \ac{AoA} of the $k^{th}$ target. We have simulated two targets, i.e. $K = 2$. Also,  $P$ is the number of Monte-carlo trials. We observe that at an \ac{MSE} of $0.2$ is achieved with the sequential \ac{NLMS} method at $\SNR = -18.55 \dB$ utilizing $M = 16$ reflective elements. The same \ac{MSE} can be achieved at $\SNR = -22.8 \dB$ with the batch \ac{NLMS} method, translating to a gain of $4.25\dB$. Comparing both methods to the no-\ac{RIS} baseline, a gain superior to $20\dB$ can be attained. This demonstrates the advantage of the proposed \ac{RIS}-aided model, as the accuracy of the \ac{AoA} estimates are drastically improved. On the other hand, since the proposed methods also offer joint detection of number of targets, we also study the probability of correct detection for the proposed methods.


In Fig. \ref{fig:PDvsSNR}, we study the probability of correct detection of the proposed methods, and compare it with the baseline scheme. The detection probability is computed as follows
\begin{equation}
	P_D = \frac{\text{no. of correct target enumeration}}{\text{no. of trials}}.
\end{equation}
Setting a level of $P_D = 0.9$, and when no \ac{RIS} is deployed, the required \ac{SNR} is $\SNR = -4.4 \dB$ for batch \ac{NLMS}, as opposed to $\SNR = 0.95 \dB$ for sequential \ac{NLMS}. A gain of $16\dB$ in terms of \ac{SNR} is noticed, when using an \ac{RIS} with only $M=16$ reflective elements. An additional $4 \dB$ gain can be attained when doubling the number of \ac{RIS} reflective elements.


To get a better sense of target localization accuracy, we study the \ac{SRP} in Fig. \ref{fig:SRPvsSNR}. To this end, the \ac{SRP} is defined as follows
\begin{equation}
	\SRP = \frac{\text{no. of trials with all } \theta_k = \hat{\theta}_k}{\text{no. of trials}}.
\end{equation}
As it can be concluded, the non-aided \ac{RIS} scenario cannot achieve a perfect \ac{SRP} with $L = 100$ time samples and $N_{\tt{PR}} = 8$ antenna elements. The \ac{SRP} is saturated in this case at $66\%$ starting at $\SNR = 16\dB$. In contrast, we observe that $66\%$ SRP can be achieved with $\SNR = -8 \dB$ with only $M = 16$ reflective elements. Furthermore, and by doubling the number of reflective elements, we can achieve an additional gain of about $18\dB$, for $\SRP = 66\%$. Additionally, the \ac{SRP} trespasses $95\%$ for $M=32$ elements when the batch \ac{NLMS} method is utilized. 

Next, we study the \ac{CDF} as a function of absolute errors in Fig. \ref{fig:CDFvsError}, i.e. $\Pr(\vert \theta - \hat{\theta} \vert < e)$. We see that the baseline approach requires an error of $e = 32^\circ$ for a probability of $0.9$ for both batch and sequential \ac{NLMS} algorithms. On the other hand, when adopting the \ac{RIS}-aided approach, we see that this error can be reduced to an order of magnitude of about $10$. Namely, we see that $\Pr(\vert \theta - \hat{\theta} \vert < 1.9) = 0.9$ for sequential \ac{NLMS} and $\Pr(\vert \theta - \hat{\theta} \vert < 1.75) = 0.9$ for batch \ac{NLMS} with $M = 16$ reflective elements. Furthermore, by doubling the number of reflective elements, the error becomes close to the resolution of $\Theta_{\tt{grid}}$.

Another important performance is the behavior of the \ac{MSE}, when varying the number of targets in the scene. Each new target introduced in the scene is spaced at $5^\circ$ further away from the previous target, measured with respect to the \ac{RIS}. Nevertheless, the \ac{SNR} at the \ac{PR} is set to $\SNR = 0 \dB$. In Fig. \ref{fig:MSEvsNoOfTargets}, we plot the \ac{MSE} as a function of number of targets. As we can observe, as soon as we have $2$ closely spaced targets, the \ac{MSE} of both methods trespasses an \ac{MSE} of $1$, when no \ac{RIS} is employed. However, by using $M = 16$ reflective elements, the number of closely spaced targets that can be resolved with an \ac{MSE} of $1$ almost quadruples. Furthermore, by doubling the number of reflective elements, we can further double the number of targets.

We also study the probability of detection of targets in Fig. \ref{fig:PDvsNoOfTargets}, for different values of number of targets. The targets are generated the same way as in the previous experiment. Moreover, the \ac{SNR} at the \ac{PR} is fixed to $\SNR = 0 \dB$. Setting a desired probability of detection to $P_D = 0.9$, we see that a no-\ac{RIS} approach cannot reach this limit, even with only $1$ target. Comparing this baseline scheme with an \ac{RIS}-aided approach with $M = 16$ reflective elements, we can see that with $3$ targets, we are at almost $P_D \simeq 0.98$ for both batch and sequential \ac{NLMS} methods. We can further double the number of resolvable targets to $6$, by doubling the number of reflective elements at the \ac{RIS} to $M = 32$. 

Furthermore, Fig. \ref{fig:PDvsAngularSeparation} reports the performance against angular separation. We fix two targets and vary the angle between them, i.e. the second target is placed at $\theta_2 = \theta_1 + \Delta$. Nevertheless, the $\SNR$ at the \ac{PR} is set to $20\dB$. In such a setting, results demonstrate that a no-\ac{RIS} approach requires an angular separation of $\Delta = 9^\circ$ to exceed a level of $P_D = 0.8$. This resolution can be improved by utilizing an \ac{RIS}-aided scenario with $M = 32$ reflective elements, where for the same $P_D$, the angular separation is improved by a factor of $2.25$, i.e. $\Delta = 4^\circ$. This resolution can even be improved towards $\Delta \simeq 2.3^\circ$ for the same $P_D$, just by doubling the amount of reflective elements.

Fig. \ref{fig:spectra} shows the spectra obtained by the proposed \ac{NLMS} filtering methods, namely Fig. \ref{fig:spectrum-LMS-batch} shows the spectrum of {\tt{\textbf{Algorithm \ref{alg:cap}}}}, whereas Fig. \ref{fig:spectrum-LMS-seq} shows that of {\tt{\textbf{Algorithm \ref{alg:cap2}}}}. It is clear from both simulations that the peaks are sharp enough to successfully resolve the $K = 4$ targets. Notice that for the sequential \ac{NLMS} filter, we remind the reader that {\tt{\textbf{Algorithm \ref{alg:cap2}}}} yields a spectrum $P(\theta)$ per epoch. Naturally, the spectrum starts off with peaks that skew from their true positions, resulting in a poor resolution. Nevertheless, with increasing epoch number, the spectrum smoothes down and peaks converge to their true positions. 
%
%

\section{Conclusions and Future Insights}
\label{sec:conclusion}
In this manuscript, target localization in terms of \ac{AoA} is accomplished via a newly proposed architecture well-suited for communication-centric \ac{ISAC} designs, consisting of an \ac{RIS} and monitored by a \ac{PR}, where target detection and estimation are performed. Towards such a setting, we posit an optimization framework enabling us to configure the phase shifters at the \ac{RIS} in order to reliably mirror the useful reflections caused by the intended targets. Furthermore, we have derived two methods, motivated by the problem of minimal information available at the disposal of the \ac{PR}. These methods are based on the \ac{NLMS} filter, and are well-suited for joint detection of number of targets and estimation of the \ac{AoAs} of the corresponding targets within the proposed \ac{RIS}-aided communication-centric \ac{ISAC} architecture. Numerical simulations validate the superior performance of both methods, when applied on the proposed architecture.

Future work will be oriented towards an adaptive beamformer, where the goal would be to learn the covariance matrix of the interference plus noise part, thanks to the target localization part, which could leverage such necessary information for covariance construction. Moreover, further extensions towards more sophisticated infrastructure, such as simultaneous AP transmissions or hybrid RIS where the goal would be to guarantee a reflected signal with good communications and radar characteristics. Furthermore, one could exploit additional sensing parameters with our proposed scenario, such as distance estimation via propagation delays and velocity estimation via doppler information. Also, one could generalize the scenario at hand toward two-dimensional and three-dimensional localization, by exploiting azimuth-elevation information. Even more, multiple \ac{RIS} could be deployed, therefore, the \ac{PR} would create multiple beams to align its look direction at those surfaces. A generalization of the aforementioned direction is to have each \ac{RIS} oriented towards possibly different objectives, i.e. sensing, communication, and security. Future research will also be shifted towards a deeper analysis oriented towards fundamental limits of the proposed model, i.e. Cram\'er-Rao bound analysis. This could allow to determine the required number of reflective \ac{RIS} elements required for a given positioning accuracy.

%
%
%

\bibliographystyle{IEEEtran}
\bibliography{refs}

\vfill

\end{document}

%% file: acronyms.tex
\begin{acronym}
    \acro{MIMO}{multiple-input, multiple-output}
    \acro{AWGN}{additive white Gaussian noise}
    \acro{KPI}{key performance indicator}
      \acro{KPIs}{key performance indicators}  
    \acro{D2D}{device-to-device}
    \acro{ISAC}{integrated sensing and communications}
    \acro{DFRC}{dual-functional radar and communication}
    \acro{AoA}{angle of arrival}
    \acro{AoAs}{angles of arrival}
    \acro{ToA}{time of arrival}
    \acro{EVM}{error vector magnitude}
    \acro{CPI}{coherent processing interval}
    \acro{eMBB}{enhanced mobile broadband}
    \acro{URLLC}{ultra-reliable low latency communications}
    \acro{mMTC}{massive machine type communications}
	\acro{QCQP}{quadratic constrained quadratic programming}
	\acro{BnB}{branch and bound}
	\acro{SER}{symbol error rate}
	\acro{LFM}{linear frequency modulation}
	\acro{BS}{base station}
	\acro{UE}{user equipment}
	\acro{DL}{deep learning}
	\acro{CS}{compressed sensing}
	\acro{PR}{passive radar}
	\acro{LMS}{least mean squares}
	\acro{NLMS}{normalized least mean squares}
	\acro{RIS}{reconfigurable intelligent surface}
	\acro{RISs}{reconfigurable intelligent surfaces}
	\acro{IRS}{intelligent reflecting surface}
	\acro{IRSs}{intelligent reflecting surfaces}
	\acro{LISA}{large intelligent surface/antennas}
	\acro{LISs}{large intelligent surfaces}
	\acro{OFDM}{orthogonal frequency-division multiplexing}
	\acro{EM}{electromagnetic}
	\acro{MSE}{mean squared error}
	\acro{SNR}{signal-to-noise ratio}
	\acro{SRP}{successful recovery probability}
	\acro{CDF}{cumulative distribution function}
	\acro{ULA}{uniform linear array}
	\acro{RSS}{received signal strength}
\end{acronym}